\documentclass[twocolumn,amsfonts,showpacs,superscriptaddress,nofootinbib]{revtex4-1}
\usepackage{pgfplots}
\usepackage{graphicx}
\usepackage{amssymb,amsmath,amsthm,booktabs,mathtools}
\usepackage{bm}
\usepackage{color}

\newcommand{\bc}{\begin{center}}
\newcommand{\ec}{\end{center}}
\def\ba#1{\begin{array}{#1}\displaystyle}
\newcommand{\ea}{\end{array}}

\newcommand{\beq}{\begin{equation}}
\newcommand{\eeq}{\end{equation}}
\newcommand{\beqa}{\begin{eqnarray}}
\newcommand{\eeqa}{\end{eqnarray}}
\newcommand{\no}{\nonumber}
\newcommand{\n}{\nonumber\\}
\newcommand{\bi}{\begin{itemize}}
\newcommand{\ei}{\end{itemize}}

\def\lt#1{\left#1}
\def\rt#1{\right#1}
\def\t#1{\tilde{#1}}

\def\b#1{\bar{#1}}
\def\frc#1#2{\frac{#1}{#2}}

\newcommand{\p}{\partial}

\newcommand{\ket}{\rangle}

\newcommand{\R}{{\mathbb{R}}}

\newcommand{\ri}{{\rm i}}
\newcommand{\dd}{{\rm d}}

\newcommand{\eff}{{\rm eff}}

\def\eqref#1{(\ref{#1})}

\newcommand{\btheta}{\bm{\theta}}
\newcommand{\balpha}{\bm{\alpha}}

\usepackage{amsmath}	
\begin{document}

\title{Soliton gases and generalized hydrodynamics}

\author{Benjamin Doyon}
\affiliation
{Department of Mathematics, King's College London, Strand, London WC2R 2LS, UK
}
\author{Takato Yoshimura}
\affiliation
{Department of Mathematics, King's College London, Strand, London WC2R 2LS, UK
}
\author{Jean-S\'ebastien Caux}
\affiliation
{Institute for Theoretical Physics Amsterdam and Delta Institute for Theoretical Physics, University of Amsterdam, Science Park 904, 1098 XH Amsterdam, The Netherlands
}
\begin{abstract}
Dynamical equations in generalized hydrodynamics (GHD), a hydrodynamic theory for integrable quantum systems at the Euler scale, take a rather simple form, even though an infinite number of conserved charges are taken into account. We show a remarkable quantum-classical equivalence: we demonstrate the equivalence between the equations of GHD, and the Euler-scale hydrodynamic equations of a new family of classical gases which generalize the gas of hard rods. In this family, the ``quasi-particles", upon colliding, jump forward or backward by a distance that depends on their velocities, generalizing the jump forward by the rods' length of the fixed-velocity tracer upon elastic collision
of two hard rods. Such velocity-dependent position shifts are characteristics of classical soliton scattering. The emerging hydrodynamics of a quantum integrable model is therefore that of the classical gas of its solitons. This provides a ``molecular dynamics'' for GHD which is numerically efficient and flexible. This is directly applicable, for instance, to the study of inhomogeneous dynamics in integrable quantum chains and in the Lieb-Liniger model realized in cold-atom experiments.
\end{abstract}

\maketitle
\noindent {\it Introduction.}\quad It is widely believed and acknowledged that the
late-time and large-scale dynamics of interacting systems, weather quantum or not, is
well described by hydrodynamics. The applicability of hydrodynamics
encompasses a large number of many-body systems, from classical gases and interacting quantum field theories \cite{spohn,jeon} where few hydrodynamic variables are necessary, to more exotic systems such as the classical hard-rod model \cite{bds}, where velocities are preserved in each collision. Recently, the realm of hydrodynamics was extended to integrable quantum models by accounting for the infinity of nontrivial conservation laws they admit \cite{ghd,bertini1}. In this context, on large (Eulerian) scales one assumes that the system reaches, locally at each points, a generalized Gibbs ensemble (GGE). The theory describing this was dubbed generalized hydrodynamics (GHD). This theory is applicable to many integrable models, including integrable quantum chains and field theory. In particular, it is applicable to the Lieb-Liniger model, and can therefore be used to describe the inhomogeneous dynamics in quasi-one-dimensional cold atom setups \cite{review_giamarchi} such as that of the celebrated quantum Newton cradle \cite{qnc}.

In this paper, we first observe that a special case of the GHD equations reproduces exactly the equations for a gas of hard rods on the line, whose dynamics is free except for elastic collisions where velocities are exchanged. The fluid equations for such a gas were mathematically derived by Boldrighini, Dobrushin, Sukhovin in 1983  \cite{bds}, and the domain wall problem studied recently  \cite{ds} paralleling the recent solution in the quantum context from GHD \cite{ghd}. We then show that a certain modification of the hard-rod dynamics leads exactly to the general form of GHD. In the modified  problem, colliding rods are replaced by point-like ``quasi-particles" which upon colliding jump, backward or forward, by a distance that depends on their velocities. These quasi-particles generalize the tracer of a given velocity in the hard rod problem, which jump forward by the fixed rods' length at rods collisions. This thus gives an explicit quantum-classical correspondence at the hydrodynamic level. The velocity-dependent spatial shift of trajectories is an effect that is known to occur in classical soliton scattering. GHD can therefore be interpreted as hydrodynamic equations for a classical soliton gas, and thus should be applicable to classical field theory problems. The spatial shift exactly agrees with that of wave-packet constructions of solitons in quantum integrable models, studied and numerically observed in \cite{vlijm}. This means that, from the viewpoint of Eulerian hydrodynamics, all quantum effects can be accounted for by considering the two-body classical-like scattering of soliton wave packets. This provides a nontrivial connection between the equations of state, or the hydrodynamics, of a quantum gas, and the classical soliton picture that underlies it. The soliton gas problem is extremely easy to implement on the computer, and thus this gives rise to a classical ``molecular dynamics" that is numerically efficient and flexible enough to be able to account for external forces and, eventually, other effects such as integrability breaking and viscosity. The classical gas picture also gives a geometric framework allowing for exact solutions to GHD (developed in \cite{DSY}).

\vspace{0.2cm}

\noindent{\it Generalized hydrodynamics.}\quad  In the context of many-body quantum physics, hydrodynamics is a theory for the dynamics of quantum averages of local observables in non-homogeneous, non-stationary states. The principle of hydrodynamics is that when profiles of local averages become smooth enough in space-time, at every point, local averages may be evaluated, to a good approximation, within an entropy-maximized homogeneous state. Local averages are thus related to each other in the same way they are in entropy-maximized states: the equations of state hold. Entropy maximization occurs with respect to the set of conservation laws afforded by the dynamics, and the associated Lagrange parameters, which determine the local state, depend on space-time. Eulerian hydrodynamics (that is, hydrodynamics neglecting viscosity effects) is obtained simply by imposing the conservation laws for local averages of densities and currents. This thus transfers quantum dynamics into a classical dynamics for the Lagrange parameters, or for any other hydrodynamic variables (parametrizations of the state).

These concepts may be applied to one-dimensional integrable models, where infinitely many local and quasi-local conserved charges are considered in entropy maximization. In this context, local averages are to be evaluated in generalized Gibbs ensembles (GGEs) (for reviews see \cite{GGEreviews}), and the hydrodynamic theory based on GGEs is referred to as generalized hydrodynamics (GHD) \cite{ghd}.  
The most convenient hydrodynamic variables of GHD are defined using the quasi-particle description of Bethe ansatz integrable models. Consider the set of quasi-particle states $|\theta_1\ldots \theta_n\ket_{a_1,\ldots, a_n}$. We parametrize each quasi-particle by its ``rapidity'' $\theta$ and specie $a$. The rapidity can be chosen as any parametrization of the two fundamental characteristics of the model, the energy $E(\btheta)$ and the momentum $p(\btheta)$, which form the group velocity $v^{\rm gr}(\btheta)=E'(\btheta)/p'(\btheta)$ (here and below we use boldface greek letters to represent a pair of spectral parameter and particle type, $\btheta=(\theta, a)$, and the prime symbol ($'$) denotes rapidity derivatives $\dd/\dd\theta$). For instance, in relativistic models, one takes $v^{\rm gr}(\btheta) = \tanh\theta$ and $p(\btheta) = m_a\sinh(\theta)$, while in Galilean models $v^{\rm gr}(\btheta) = \theta$ and $p(\btheta) = m_a\theta$. In integrable models, the interaction can be fully characterized by the two-particle scattering amplitude $S(\btheta,\balpha)$, and we define the differential scattering phase, $\varphi(\btheta,\balpha) = -\ri\, \dd S(\btheta,\balpha)/\dd\theta$. We assume $\varphi(\btheta,\balpha)$ to be symmetric.

In the thermodynamic limit, one represents a GGE state by its quasi-particle density $\rho_{\rm p}(\btheta)$; the number of quasi-particles of type $a$ in the phase space element $[x,x+\dd x]\times [p(\btheta),p(\btheta)+p'(\btheta)\dd\theta)]$ is $\rho_{\rm p}(\btheta)\dd\theta\dd x$. This quantity fully determines the GGE state, and any local average can be evaluated in terms of $\rho_{\rm p}(\btheta)$.
Within this picture, one also introduces the density of states $\rho_{\rm s}(\btheta)$ defined by $2\pi \rho_{\rm s}(\btheta) = p'(\btheta) + \int \dd\balpha\, \varphi(\btheta,\balpha) \rho_{\rm p}(\balpha)$, and the occupation function characterizing the proportion of available states that are occupied by quasi-particles, $n(\btheta) = \rho_{\rm p}(\btheta)/\rho_{\rm s}(\btheta)$.

It was shown in \cite{ghd,bertini1} that the infinity of hydrodynamic conservation laws of GHD lead to the following equivalent continuity and convective equations:
\begin{align}\label{cont1}
	\p_t \rho_{\rm p}(\btheta) + \p_x(v^{\rm eff}(\btheta)\rho_{\rm p}(\btheta)) &=0\\ \label{cont2}
	\p_t n(\btheta) + v^{\rm eff}(\btheta)\p_xn(\btheta) 
	&= 0
\end{align}
where the {\em effective velocity} $v^{\eff}(\btheta)$ solves the equation
\beq\label{veos}
	v^{\rm eff}(\btheta) = v^{\rm gr}(\btheta)
	+\int \dd\balpha\,\frc{\varphi(\btheta,\balpha)}{p'(\btheta)}\,
	\rho_{\rm p}(\balpha)\,
	(v^{\rm eff}(\balpha)-v^{\rm eff}(\btheta))
\eeq
(here and below $\int \dd\btheta = \sum_a \int_\R \dd \theta$, and we suppress the explicit $x,t$ dependence for lightness of notation). The effective velocity $v^{\eff}(\btheta)$ \cite{bonnes,ghd,bertini1} serves as a velocity of the quasi-particle $\btheta$ in the fluid cell at $(x,t)$ (microscopically, it is the velocity of the elementary excitation carrying the state $\btheta$). The quasi-particle spectral density $\rho_{\rm p}(\btheta)$ is a conserved fluid density, while $n(\btheta)$ is a convected fluid variable. In particular, one may evaluate averages of any local conserved density $q$ and its current $j$ via \cite{ghd,bertini1}
\beq\label{kern}
	{\tt q} = \int \dd\btheta\,
	\rho_{\rm p}(\btheta) h(\btheta),\quad
	{\tt j} = \int \dd\btheta\,
	v^{\rm eff}(\btheta)\rho_{\rm p}(\btheta) h(\btheta)
\eeq
where $h(\btheta)$ is the quantity of charge carried by the quasi-particle $\btheta$ (the one-particle eigenvalue of the associated conserved charge).

For instance, in the Lieb-Liniger (LL) model \cite{LLmodel}, a Galilean invariant model for interacting Bose gases experimentally realizable \cite{review_giamarchi}, there is a single particle specie in the repulsive regime, and the differential scattering phase takes the form
\beq\label{ll}
	\varphi(\theta,\alpha) = 2c/(\theta^2+c^2) \qquad\mbox{(Lieb-Liniger)}
\eeq
where $c$ is the coupling strength (see the Supplementary Material (SM) for a description of the LL model in the attractive regime). The observable for the number of particles in the gas corresponds, in the above formulae, to the choice $h(\btheta)=1$. In the LL model, Eqs. \eqref{cont1}-\eqref{kern} were derived in \cite{ghd}, and in the XXZ quantum spin chains, they were first obtained in \cite{bertini1}.

GHD was also considered for evolution within inhomogeneous fields, such as force or temperature fields \cite{dy}. We may represent such a field by simply taking the energy to be space-dependent $E(\btheta) = E(x;\btheta)$. Then the following equivalent equations hold:
\begin{align}\label{cont1force}
	\p_t \rho_{\rm p}(\btheta) + \p_x(v^{\rm eff}(\btheta)\rho_{\rm p}(\btheta)) + \p_\theta (a^{\rm eff}(\btheta) \rho_{\rm p}(\btheta))&=0\\ \label{cont2force}
	\p_t n(\btheta) + v^{\rm eff}(\btheta)\p_xn(\btheta) 
	+ a^{\rm eff}(\btheta)  \p_\theta n(\btheta)&= 0
\end{align}
where $v^{\rm eff}(\btheta)$ is still given by \eqref{veos}, but with space-dependent group velocity. The effective acceleration is the time derivative $a^{\rm eff}(\btheta)=\dd v^{\rm eff}(\btheta)/\dd t|_{\rho_{\rm p}}$
under the time change of the group velocity due to the force $F(\btheta) = -\p_x E(\btheta)$ at constant fluid state $\rho_{\rm p}$, obtained from differentiating \eqref{veos} with $\dd v^{\rm gr}(\btheta)/ \dd t = F(\btheta)/p'(\btheta)$. It simplifies to the ordinary accleration $a^{\rm eff}(\btheta) = F(\btheta)/p'(\btheta)$ when the latter quantity is independent of $\btheta$. In the single-specie Galilean case with a force potential $V$, such as in the repulsive LL model, we have $E(\theta)=m\theta^2/2+V$ and $a^{\rm eff}(\theta) = -\p_x V/m$. See \cite{dy} for details.

\vspace{0.2cm}

\noindent {\it Classical gases.}\quad Let us first recall the classical hard rod model \cite{spohn,bds}. Rods (non-intersecting one-dimensional segments) of a fixed length $d$ move inertially at various velocities $v$ on the infinite line, except for elastic collisions at which they exchange their velocities. The emergence of hydrodynamic equations on large scales in this model for a large class of initial conditions was rigorously demonstrated \cite{bds}. Let $\rho_{\rm cl}(v)$ be the density of rods with velocity $v$ (that is, $\rho_{\rm cl}(v)\,\dd x \,\dd v$ is the number of rod centers lying within the phase space element $[x,x+\dd x]\times [v,v+\dd v]$). The hydrodynamic equations read
\begin{equation}
 \p_t\rho_{\rm cl}(v)+\p_x(v^{\rm eff}_{\rm cl}(v)\rho_{\rm cl}(v))=0,
\end{equation}
where $v^{\rm eff}_{\rm cl}(v)$ satisfies \cite{bds}
\begin{equation}\label{vhard}
 v^{\rm eff}_{\rm cl}(v)=v+d\int\dd w\,\rho_{\rm cl}(w)(v^{\rm eff}_{\rm cl}(v)-v^{\rm eff}_{\rm cl}(w)).
\end{equation}
These are exactly the same equations as \eqref{veos} and \eqref{cont1}, in the Galilean case ($v=\theta$),  without inhomogeneous fields, with a single unit-mass particle specie, and with the identifications $d=-\varphi(v,w)$, $\rho_{\rm p}(v)=\rho_{\rm cl}(v)$, and $v^{\rm eff}(v)=v^{\rm eff}_{\rm cl}(v)$. In the quantum model, this corresponds to a purely exponential scattering phase, $S(\theta,\alpha) = e^{-\ri d\,(\theta-\alpha)}$. This is the non-relativistic limit of a scattering phase which was considered in \cite{GL} in a different context.

The above observation suggests that if we allow the rods to collide more ``softly'', in such a way that $d$ becomes velocity dependent, the hydrodynamics of the emerging gas might be identical to that of GHD.

Rod lengths are distances at which two rod centers exchange their velocities. One could attempt to directly extend this to velocity dependent rod lengths: neighboring rods with velocities $v$ (left) and $w$ (right), with $v>w$, exchange their velocities when their centers are at distance $d(v,w)$ (as if rods were elastically contractible instead of hard). However this causes difficulties with respect to many-body scattering: after the instantaneous exchange of velocities, other rods in the immediate neighborhood might be nearer than the velocity-dependent minimal distance. It also causes difficulties for negative lengths. Instead, let us modify the microscopic dynamics to one that is more easily generalizable, keeping the same hydrodynamics.

Consider the trajectory of a velocity-tracer, following the center of the rod carrying velocity $v$. Its trajectory is that of a free particle, except for jumps by a distance $d$ when rod collisions occur. The tracer is a quasi-particle of velocity $v$. In this language, quasi-particles are point-like objects, and two quasi-particles at positions $x_1<x_2$ ``collide" if their positions satisfy $x_2-x_1=d$ and their velocities $v_1>v_2$, at which instant $x_1\mapsto x_1+d$ and $x_2\mapsto x_2-d$, and quasi-particle trajectories cross. The important property of this microscopic dynamics is that every crossing of two quasi-particles' trajectories comes with such a shift by $d$, in microscopic time (here, instantaneously). As is clear from the derivation below, any microscopic dynamics with this property will lead to the same hydrodynamics.

We may therefore modify this classical gas by proclaiming collisions to occur at $x_2=x_1$ instead. At a collision, the involved quasi-particles instantaneously jump, like fleas, by a distance $d$ as above. The jump is ``forward'': the quasi-particle on the left (right) jumps towards the right (left). This classical ``flea gas" is generalized easily: a collision between quasi-particles of velocities $v$ and $w$ occasion a jump by $d(v,w)$, forward if positive, backward if negative. A jump should be seen as an infinitely-fast displacement, during which more collisions can occur, occasioning new jumps in a chain reaction that re-organizes the quasi-particles' positions in the local neighborhood. See the SM for the precise flea-gas algorithm and other microscopic procedures.

We now argue that this reproduces GHD. We are looking for the effective velocity $v^{\rm eff}_{\rm cl}(v)$ of a test quasi-particle, defined through the actual distance that it travels in a macroscopic time $\Delta t$,
\beq\label{e1}
	\Delta x = \Delta t\, v^{\rm eff}_{\rm cl}(v).
\eeq
The gas is characterized by the density $\rho_{\rm cl}(v)$, and by standard arguments, the continuity equation holds, $\p_t \rho_{\rm cl}(v) + \p_x(v^{\rm eff}_{\rm cl}(v)\rho_{\rm cl}(v))=0$. The quantity $\Delta x$ results from the linear displacements at velocity $v$, given by $\Delta t\, v$, along with the accumulation of jumps the quasi-particle undergoes as it travels through the gas. The oriented distance jumped due to hitting a quasi-particle that has velocity $w$ is ${\rm sign}(v-w)d(v,w)$. The average number of quasi-particles of
velocity between $w$ and $w+\dd w$ that has been crossed, is the total number $\dd w\,\rho_{\rm cl}(w)\Delta x$ present within the length $\Delta x$, times the probability $\Delta t/\Delta x \times |v^{\rm eff}_{\rm cl}(v)-v^{\rm eff}_{\rm cl}(w)|$ that the test particle crosses such a quasi-particles in time $\Delta t$. Assuming that the effective velocity is monotonic with $v$ (see the SM), the total jumped distance is therefore
\beq\label{e2}
	\int \dd w\,d(v,w)\,\rho_{\rm cl}(w)\,\Delta t\,
	(v^{\rm eff}_{\rm cl}(v)-v^{\rm eff}_{\rm cl}(w)).
\eeq
Equating \eqref{e1} with the sum of $\Delta t\,v$ and \eqref{e2} we find
\beq\label{veffrod}
	v^{\rm eff}_{\rm cl}(v)
	= v + \int \dd w\,d(v,w)\,\rho_{\rm cl}(w)\,
	(v^{\rm eff}_{\rm cl}(v)-v^{\rm eff}_{\rm cl}(w)).
\eeq

It is clear that the GHD equations \eqref{cont1}, in the case of a single specie, reproduce the hydrodynamics of the above classical gas under the following identification:
\beq\label{d1}
	\rho_{\rm cl}(v)\dd v = \rho_{\rm p}(\theta)\dd\theta,\quad
	v = v^{\rm gr}(\theta),\quad v^{\rm eff}_{\rm cl}(v) = v^{\rm eff}(\theta)
\eeq
along with
\beq\label{d2}
	d(v,w) = -\varphi(\theta,\alpha)/p'(\theta).
\eeq
The above derivation is readily generalizable to many species, with, in \eqref{veffrod}, the velocity parameters $v,w$ replaced by doublets $\bm{v} = (v,a),\,\bm{w} = (w,b)$, and the driving  velocity value $v$ replaced by the appropriate value $v^{\rm gr}(\bm{v})$ for specie $a$. We recover \eqref{veos} by reparametrization. It is also clear that \eqref{cont1force} holds in the single-specie case if an external force field is applied, as the acceleration times $\rho_{\rm cl}(\btheta)$ directly gives the current of quasi-particles in velocity space. We leave the general case of \eqref{cont1force} for a future work.

\vspace{0.2cm}

\noindent {\it  The quantum-classical dictionary.}\quad The above shows that GHD, derived in quantum integrable models using quantum integrability, also describes the hydrodynamic limit of {\em classical} models of integrable dynamics: the hard rod gas and the flea gas generalization. Eqs. \eqref{d1} and \eqref{d2} give a quantum-classical dictionary. Further elements of the dictionary are as follows. Consider the free space fraction (in the hard rod gas, the fraction of a unit length where there is no rod at all; in the general case, that available omitting the distances jumped if forward, or adding them if backward):
\beq
	\rho_{\rm free}({\bm v}) = 1-\int \dd {\bm w} \, d({\bm v},{\bm w})\rho_{\rm cl}({\bm w}).
\eeq
We recognize this quantity as, up to a factor, the {\em density of state} in the quantum problem, 
\beq
	\rho_{\rm free}({\bm v}) = 2\pi \rho_{\rm s}(\btheta)/p'(\btheta).
\eeq
For the occupation function $n(\btheta)$, we then have
\beq
	\frc{\rho_{\rm cl}({\bm v})\dd v\,}{\rho_{\rm free}({\bm v})} = n(\btheta)\frc{p'(\btheta) \dd \theta}{2\pi}.
\eeq
The left-hand side is the classical number of quasi-particles per unit length of free space. Note that for negative jumped distances, the ``fraction" of a unit length that is free is greater than one: the effect of quasi-particle scattering is to increase the space available.

The classical picture provides a clear interpretation of the effective velocity. Let us write it in the form
\beq\label{veff}
	v^{\rm eff}_{\rm cl}(\bm{v}) = \frc{v^{\rm gr}(\bm{v})
	 - \int \dd \bm{w}\,
	d(\bm{v},\bm{w})\rho_{\rm cl}(\bm{w}) v^{\rm eff}_{\rm cl}(\bm{w})}{
	1 - \int \dd \bm{w}\,
	d(\bm{v},\bm{w})\rho_{\rm cl}(\bm{w})}.
\eeq
Suppose $d(v,w)<0$. In this case, the gas {\em slows down} a test quasi-particle, as it is affected by backwards jumps at collisions. The denominator accounts for this friction effect. If the gas also carries a drift, then the test particle is carried by the gas itself for the same reason. The second term in the numerator accounts for this. These two competing effects were noticed in \cite{ghd} when studying the steady state emerging from by a domain-wall initial condition in the sinh-Gordon and Lieb-Liniger models. In the case $d(v,w)>0$, the opposite occurs.

Finally, it is worth noting that in the classical problem, the rod length $d(v,w)$ can be chosen at will, and in particular the full dynamics is invariant under simultaneous scaling of space, time and rod lengths. In the quantum case, there is no such freedom: the differential scattering phase $\varphi(\btheta,\balpha)$ is fixed by the scattering of quasi-particles (the a fundamental length scale is determined by $\hbar$).

\vspace{0.2cm}

{\it Soliton gases.}\quad The above intriguing quantum-classical correspondence might be explained in terms of soliton gases. In classical soliton scattering, two solitons retain, asymptotically, their form and their speeds, the only change being in shifts of their trajectories. These shifts are velocity dependent, and thus, up to microscopic differences that do not affect the large-scale hydrodynamics, the flea-gas dynamics is equivalent to classical soliton scattering. Repulsive (attractive) soliton scattering correspond to positive (negative) position shifts, in agreement with the intuition from the flea gas. It is tempting to propose that GHD be the hydrodynamics arising in models of classical integrable field theory possessing soliton modes.

GHD, however, was derived in {\em quantum} integrable systems. In this context, quasi-particles are often interpreted as the quantization of classical solitons. This interpretation was recently numerically observed by forming wave packets of quasi-particle excitations in the Heisenberg quantum chain \cite{vlijm}. It was confirmed that the trajectory shifts are given by the differential scattering phase of the quantum model, in agreement with \eqref{d2}. Therefore, we conclude that at the Euler scale, quantum gases can be seen as the gas of their classical soliton wave packets. This is in agreement with the picture according to which in quantum models, any multi-particle scattering process may be seen as a sequence of well separated two-body scattering processes (at the basis of the (generalized) thermodynamic Bethe ansatz \cite{GTBA}).
\begin{figure}
\begin{tikzpicture}[scale=0.55] \begin{axis}[
title={\LARGE (a) $v^{\rm eff}$}, xlabel={\LARGE $v$}, ylabel=,xmin=-2,xmax=2,ymin=-1.5,ymax=1.5]
\addplot[blue] table {testveff10-testpart.dat};
\addplot[red] table {testveff10-leftveff-0.dat};
\end{axis} \end{tikzpicture}
\begin{tikzpicture}[scale=0.55] \begin{axis}[
title={\LARGE (b) $\rho$}, xlabel={\LARGE $x$}, ylabel=,xmin=-500,xmax=500]
\addplot[blue, mark = *, mark size = 1.5pt, only marks] table {dm2-1-density.dat};
\addplot[red] table {dm2-1-density-exact.dat};
\addplot[blue, mark = square*, mark size = 1pt, only marks] table {dm2-3-density.dat};
\addplot[red] table {dm2-3-density-exact.dat};
\addplot[blue, mark = triangle*, mark size = 1.5pt, only marks] table {dm2-5-density.dat};
\addplot[red] table {dm2-5-density-exact.dat};
\addplot[blue, mark = diamond*, mark size = 1.5pt, only marks] table {dm2-7-density.dat};
\addplot[red] table {dm2-7-density-exact.dat};
\legend{{\Large $t=100$},,{\Large $t=300$},,{\Large $t=500$},,{\Large $t=700$},}
\end{axis} \end{tikzpicture}
\caption{GHD for the LL model with $m=1$, $c=1$ is simulated using the classical flea gas. (a) Truncated Gaussian distribution $\rho_{\rm cl}(v) = 0.5e^{-v^2}\chi(-3<v<3)$. Effective velocity evaluated using approx. 1500 trajectories over a time of 1200 (blue); using the formula \eqref{veff} (red). (b) Density profile from domain wall initial condition, initial left and right temperatures $10$ and $1/3$ (resp.), at times $t=100,\,300,\,500,\,700$. Simulation with approx. 2400 quasi-particles (initial baths of lengths $1000$, open boundary condition) averaged over 1000 samples (blue); exact self-similar solution (red).}
\label{fig1}
\end{figure}
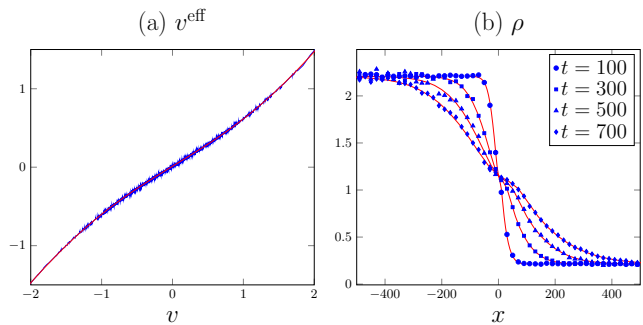

\vspace{0.2cm}
\noindent {\it Numerical checks.}\quad We have numerically simulated the classical gas corresponding to the LL model \eqref{ll} with $c=1$, $m=1$,
without force field. We have verified the form of the effective velocity by evaluating explicitly, in a homogeneous stationary gas, the total displacement of a test quasi-particle divided by the time spent, and comparing with the result of solving numerically the integral equation \eqref{veff}. See Fig. \ref{fig1}a; the agreement is excellent. This is well within the strongly interacting regime, with LL coupling parameter $\gamma = mc\rho^{-1}\approx 1.1$ ($\rho=\int \dd v\,\rho_{\rm cl}(v)$ is the particle density). We have also implemented a domain wall initial condition in the LL model, and checked that its dynamics reproduces the self-similar solution derived in \cite{ghd}. See Fig. \ref{fig1}b, as well as Fig. \ref{fig2} in the SM.

\vspace{0.2cm}
\noindent{\it Conclusion.}\quad We have developed a classical gas dynamics that reproduces, at the Euler scale, the equations of GHD for arbitrary differential scattering phase. This gives an efficient way of simulating full space-time dependent profiles solving GHD including with force fields, complementing exact solutions found in \cite{BVKM,DSY}. It is applicable to experimentally relevant models such as the LL model from any state and in any trap potential, or quantum spin chains. It also shows that the GHD equations describe classical soliton gases, and it provides a clear interpretation of friction and drag effects.  In the present formulation, scattering must be diagonalized before it is classically implemented; however perhaps non-diagonal scattering could be directly implemented in the classical gas. It would be interesting to analyze the large-deviation theory of these new classical gases, especially their relation with that of quantum problems, as well as classical gases reproducing critical quantum models (see \cite{bdreview}).

Our observations may have far-reaching implications. Soliton gases may be seen as wide generalizations of the semiclassical picture proposed in \cite{SY}, and may lead to efficient ways of evaluating correlations in certain regimes. Viscosity or other higher-derivative effects in the quantum problems will have many sources, including the finite scattering length taken into account by the classical gas, but also soliton wave packet spreading. By appropriately modifying the classical algorithm, it might be possible to phenomenologically account for such corrections to GHD, as well as for integrability-breaking processes, which would otherwise be extremely difficult to numerically implement.

\vspace{0.2cm}
\noindent{\it Acknowledgment.}\quad We thank Robert Konik and Herbert Spohn for useful discussions. Hospitality is acknowledged as follows: all authors thank SISSA and ICTP, Trieste (visits and workshop ``Entanglement and non-equilibrium physics of pure and disordered systems", July 2016), and BD and TY thank City University New York (workshop ``Dynamics and hydrodynamics of certain quantum matter", March 2017). TY is grateful for the support from the Takenaka Scholarship Foundation, and thanks Tokyo Institute of Technology for hospitality.

\newpage


\bc \Large\bf Supplementary Material
\ec

\section{Flea gas algorithm}

\begin{figure}[ht]
	\begin{center}
    	\includegraphics[width=8.0cm]{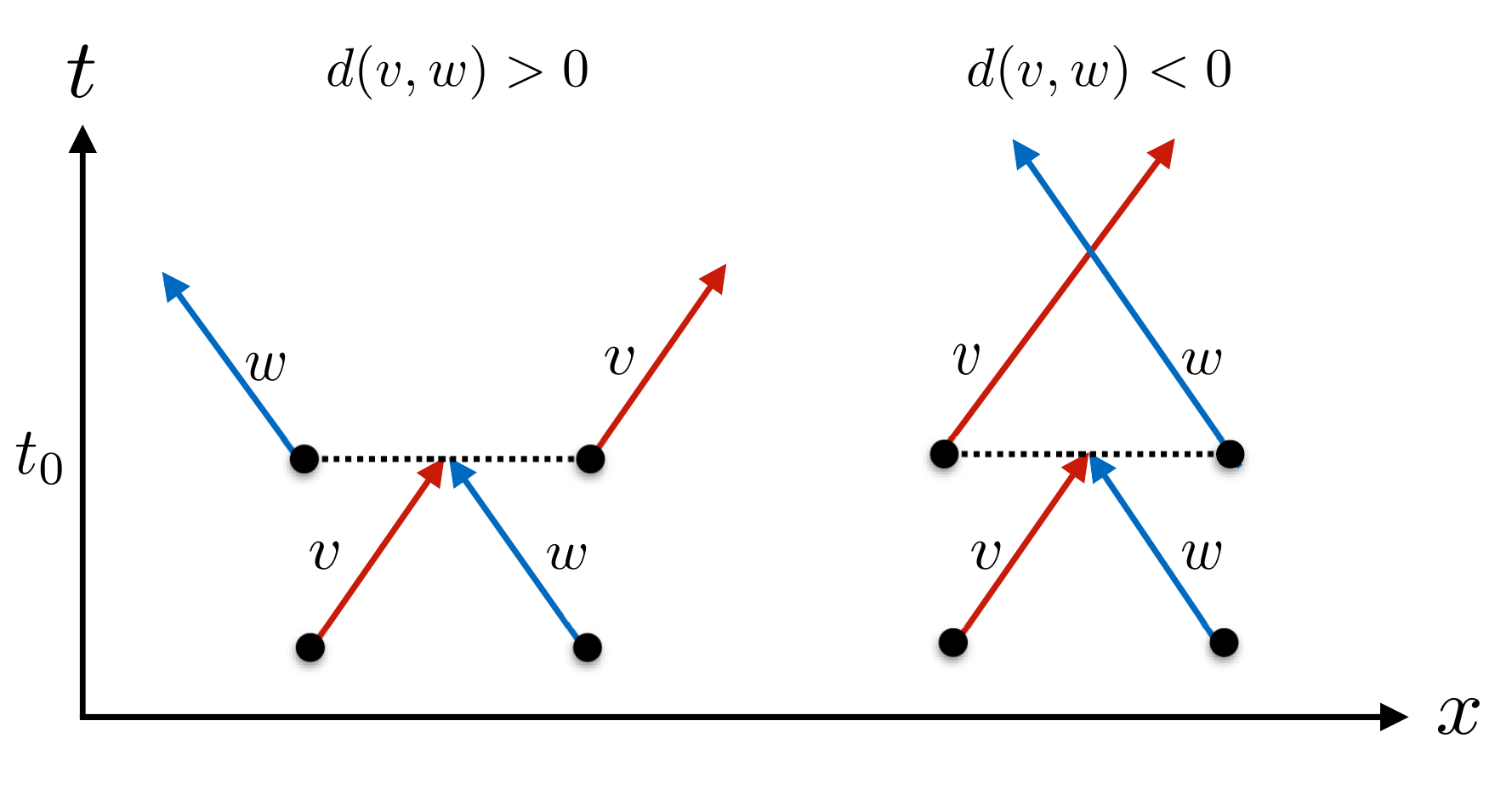}
        \caption{A cartoon picture for  collisions between quasi-particles with velocities $v$ and $w$. Upon a collision at time $t_0$, they move forward or backward depending on the sign of $d(v,w)$. }
       \label{f2}
    \end{center}
\end{figure}

The gas is represented by a chain of cells, each cell representing a quasi-particle $A$ and containing all necessary information pertaining to it (its velocity, its type, its position $A.{\tt pos}$ on the line), ordered from left to right. The procedure {\tt distance}$(A,B)$ returns the oriented distance of the jump for a collision of $A$ against $B$ (it depends on the velocities and types of $A$ and $B$); it was denoted $d(\cdot,\cdot)$ in the text, and it is positive for forward jumps, negative for backward jumps. In order not to perform a given jump, associated to a give crossing, twice, we need to mark the pairs of quasi-particle at their first collision; marked pairs, if they collide another time, just past through each other. This is of course essential if jumps are backwards, and in general has the effect that under the re-organization of quasi-particles' positions in a neighborhood of a collision, the exact distance $d(\cdot,\cdot)$ has been jumped by every quasi-particle affected by a collision. A picture for possible collisions between quasi-particles with velocities $v$ and $w$ are depicted in the Fig.\ref{f2}, and a precise algorithm for the flea gas is as follows.

\vspace{0.2cm}
{\tt
\noindent procedure Evolve:\\
Displace all until next collision;\\ Collide left-particle against right-particle;\\
repeat until evolution time has elapsed;\\
end.

\vspace{0.2cm}
\noindent procedure Collide $A$ against $B$:\\
if marked $(A,B)$:\\
\indent Exchange $A$, $B$ in the chain\\
else:\\
\indent Mark $(A,B)$;\\
\indent Jump $A$ by distance$(A,B)$;\\
\indent Jump $B$ by $-$distance$(A,B)$;\\
end.

\vspace{0.2cm}
\noindent procedure Jump $A$ by $D$:\\
if $D<0$ then side is left;\\
else side is right;\\
repeat:\\
\indent $B$ := quasi-particle to the side of $A$;\\
\indent if $B$ exists and $|A.{\tt pos} - B.{\tt pos}|<|D|$:\\
\indent\indent $D := D - B.{\tt pos}+A.{\tt pos}$;\\
\indent\indent $A$.pos := $B$.pos;\\
\indent\indent if side is left: Collide $B$ against $A$;\\
\indent\indent else: Collide $A$ against $B$;\\
\indent else:\\
\indent\indent $A$.pos := $A$.pos + $D$;\\
\indent\indent break;\\
end.
}

Above, {\tt side} refers to the side within the chain. Note that, because of the recursive process (the chain reaction of collisions), the event {\tt Jump} $A$ {\tt by distance}$(A,B)$ may also affect the position of $B$. This is fine, as long as $B$ then jumps by the correct distance afterwards. Of course, the choice of making $A$ jump
before $B$ in the {\tt Collide} procedure is arbitrary. Different orderings lead to different microscopic re-organizations
of quasi-particles' positions at collisions, but to the same large-scale Euler hydrodynamics.

An acceleration due to an external field is implicitly implemented with the {\tt Displace} procedure. There, the evolution of each quasi-particle changes, in general, both its position and its velocity, as per the usual physical laws for particles  within external force fields $F(\btheta)$.

Other algorithms are possible. In particular, it is not necessary to perform jumps ``instantaneously'' at collisions; as long as the distance $d(\cdot,\cdot)$ is added to the trajectory in a microscopic time, the large-scale effect is the same. For instance, one may associate to each quasi-particle a ``ghost velocity change'' $\Delta v$, used for time evolution in order to add an appropriate distance, but not for the calculation of the jump distance (which uses the original spectral parameter and quasi-particle type). We choose an overall time $t_{\rm micro}>0$. A quasi-particle either is in a ``ghost" state ($\Delta v>0$), or not ($\Delta v=0$). We note the time $t_0$ of the start of a ghost state. Upon a collision of $A$ with $B$, we do $\Delta v := \Delta v + \Delta t/{\tt distance}(A,B)$, so that within a time $\Delta t$, the correct distance is added. When entering a ghost state (at a collision), we set $\Delta t = t_{\rm micro}$, while for further collisions during the ghost state, we choose $\Delta t = t_{\rm micro}-t+t_0$, where $t$ is the time. The particle reverts to its normal state when $t-t_0 = t_{\rm micro}$.

\section{The Lieb-Liniger model in the attractive regime}

In first-quantized form, the Lieb-Liniger (LL) model is described by the Hamiltonian
\beq
	H = -\frc1{2m}\sum_{j=1}^N\frc{\p^2}{\p x_j^2} + c\sum_{j_1<j_2}\delta(x_{j_1}-x_{j_2}).
\eeq
With repulsive interaction, $c>0$, the spectrum of Bethe excitations is composed of a single particles, and the differential scattering phase given by \eqref{ll}. The case of attractive interaction $c=-2\b c<0$ is however not simply obtained by replacing $c$ by $-2\b c$ in \eqref{ll}. Instead, the spectrum of Bethe excitations is more complicated: it is composed of an infinity of Bethe strings, one for every positive length $j=1,2,3,\ldots$. The attractive LL model is therefore classically implemented as a gas with infinitely-many species, representing the infinitely-many string lengths,  all interacting with each other via a string-length- and velocity-dependent shift.

A thermodynamic Bethe ansatz analysis gives the following (we take $m=1/2$ for simplicity). The energy and momentum functions, for a string of length $a$ with ``pseudo-momentum" $\lambda$, are given by
\beq
	E(\lambda,j) = j\lambda^2 - \frc{\b c^2}{12} j(j^2-1),\quad
	p(\lambda,j) = j\lambda.
\eeq
The velocity is therefore
\beq
	v^{\rm gr}(\lambda,j) = 2\lambda
\eeq
which allows us to parametrize energy and momentum functions in terms of velocities,
\beq
	E({\bm v}) = \frc{jv^2}2 - \frc{\b c^2}{12} j(j^2-1),\quad
	p({\bm v}) = \frc{jv}2.
\eeq
The differential scattering phase, expressed in the pseudo-momentum coordinates $\lambda$, is the following function:
\beqa
	\lefteqn{\varphi((\lambda,j),(\lambda',j'))} && \n
	&=& (1-\delta_{j,j'})a_{|j-j'|}(\lambda-\lambda') + 2a_{|j-j'|+2}(\lambda-\lambda') + \ldots \n
	&& \ldots +2a_{j+j'-2}(\lambda-\lambda') + a_{j+j'}(\lambda-\lambda')
\eeqa
where
\beq
	a_j(\lambda) = \frc{j\b c}{\lambda^2 + (j\b c/2)^2}.
\eeq
The above can be directly applied to the formulae presented in the main text, by taking pseudo-momenta $\lambda$ as ``rapidities" (this is a good definition as the differential scattering phase depends on differences of pseudo-momenta), and identifying gas species $a$ with string lengths $j$.

\section{Monotonicity of the effective velocity}

Here we provide a demonstration that, under certain conditions on the differential scattering phase, the effective velocity is monotonic with $\theta$ (this is a fact that is used in the derivation in the main text).

Consider the case of a single particle in the spectrum:
\beq
	v^{\rm eff}(\theta) = v^{\rm gr}(\theta)
	+\int \dd\alpha\,\t\varphi(\theta,\alpha)\,
	\rho_{\rm p}(\alpha)\,
	(v^{\rm eff}(\alpha)-v^{\rm eff}(\theta)).
\eeq
where $\t\varphi(\theta,\alpha) = \frc{\varphi(\theta,\alpha)}{p'(\theta)}$. Assume first that $\t\varphi(\theta,\alpha)$ is positive (and also that $\rho_{\rm p}(\theta)$ is positive) and that its derivative has the following sign (with $' = \dd/\dd\theta$):
\beq\label{c}
	(\theta-\alpha)\t\varphi'(\theta,\alpha)\leq0
\eeq
Assume also that $\theta$ parametrizes the group velocity in a monotonic fashion, $(v^{\rm gr})'(\theta)>0$. Taking a  derivative, we find
\beqa
	\lefteqn{(v^{\rm eff})'(\theta) =(v^{\rm gr})'(\theta)\,+} \qquad   && \n &&
	+\,\int \dd\alpha\,\t\varphi'(\theta,\alpha)\,
	\rho_{\rm p}(\alpha)\,
	(v^{\rm eff}(\alpha)-v^{\rm eff}(\theta))\, -\n &&
	-\,(v^{\rm eff})'(\theta)\,\int \dd\alpha\,\t\varphi(\theta,\alpha)\,
	\rho_{\rm p}(\alpha)
\eeqa
and thus
\beqa
	\lefteqn{\lt(1+\int \dd\alpha\,\t\varphi(\theta,\alpha)\,
	\rho_{\rm p}(\alpha)\rt)
	(v^{\rm eff})'(\theta)}\qquad &&\n &=& (v^{\rm gr})'(\theta)
	+\int \dd\alpha\,\t\varphi'(\theta,\alpha)\,
	\rho_{\rm p}(\alpha)\,
	(v^{\rm eff}(\alpha)-v^{\rm eff}(\theta))\no
\eeqa
Thanks to \eqref{c}, the right-hand side of the latter equation is positive if $(v^{\rm eff})'(\theta)\geq 0$ for all $\theta$, and thus this latter condition is consistent with this integral equation. That is, if we assume that the solution is obtainable by recursion, starting with $(v^{\rm eff})'(\theta) = (v^{\rm gr})'(\theta)$, then the solution is monotonic. With decaying asymptotic for the particle density $\rho_{\rm p}(\alpha)$ it is easy to see that $v^{\rm eff}(\theta)\sim v^{\rm gr}(\theta)$ as $\theta\to\pm\infty$, and thus $v^{\rm eff}(\theta)$ covers full range of velocities.

All conditions, including \eqref{c}, are satisfied in the Lieb-Liniger model and the sinh-Gordon model.

The condition on the sign of $\t\varphi(\theta,\alpha)$ may be relaxed, as long as 
\beq
1+\int \dd\alpha\,\t\varphi(\theta,\alpha)\,
	\rho_{\rm p}(\alpha)>0
\eeq
for all $\theta$.

\section{Additional numerical checks}

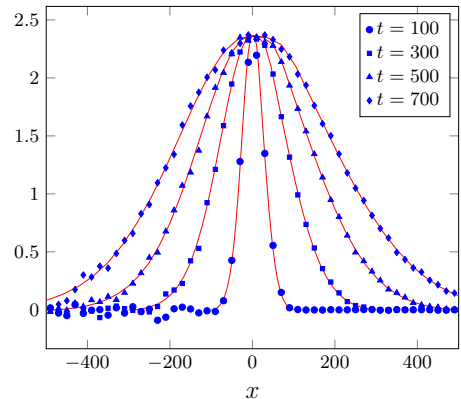
\begin{figure}
\begin{tikzpicture}[scale=0.8] \begin{axis}[
title={\ }, xlabel={\large $x$}, ylabel=,xmin=-500,xmax=500]
\addplot[blue, mark = *, mark size = 1.5pt, only marks] table {dm2-1-current.dat};
\addplot[red] table {dm2-1-current-exact.dat};
\addplot[blue, mark = square*, mark size = 1pt, only marks] table {dm2-3-current.dat};
\addplot[red] table {dm2-3-current-exact.dat};
\addplot[blue, mark = triangle*, mark size = 1.5pt, only marks] table {dm2-5-current.dat};
\addplot[red] table {dm2-5-current-exact.dat};
\addplot[blue, mark = diamond*, mark size = 1.5pt, only marks] table {dm2-7-current.dat};
\addplot[red] table {dm2-7-current-exact.dat};
\legend{{$t=100$},,{$t=300$},,{$t=500$},,{$t=700$},}
\end{axis} \end{tikzpicture}
\caption{GHD for the LL model with $m=1$, $c=1$ is simulated using the classical flea gas. Current profile from domain wall initial condition, initial left and right temperatures $10$ and $1/3$ (resp.), at times $t=100,\,300,\,500,\,700$. Simulation with approx. 2400 quasi-particles (initial baths of lengths $1000$, open boundary condition) averaged over 1000 samples (blue); exact self-similar solution (red).}
\label{fig2}
\end{figure}

In addition to the density, we have also verified numerically the accuracy of the current simulated by the classical flea gas 
as compared to the exact self-similar solution, in the Lieb-Liniger model. See Fig. \ref{fig2}. Again, the agreement is excellent, confirming the adequacy of the classical gas for representing GHD.

\end{document}